\documentstyle[aps,twocolumn,epsfig]{revtex}

\begin{document}

\title{Monte Carlo Study of Ferromagnetism in (III,Mn)V Semiconductors}

\author{John Schliemann,$^{a,b}$ J\"urgen K\"onig,$^{a,b,c}$ and 
  A. H. MacDonald$^{a,b}$}

\address{$^a$Department of Physics, The University of Texas, Austin, TX 78712}

\address{$^b$Department of Physics, Indiana University, Bloomington, IN 47405}

\address{$^c$Institut f\"ur Theoretische Festk\"orperphysik, Universit\"at
Karlsruhe, 76128 Karlsruhe, Germany}
\date{\today}

\maketitle

\begin{abstract} 

We report on Monte Carlo studies of the kinetic exchange model for 
(III,Mn)V ferromagnetic semiconductors in which $S=5/2$ local moments,
representing ${\rm Mn}^{2+}$ ions, are exchange coupled to band electrons.
We treat the ${\rm Mn}^{2+}$ spin orientations as classical degrees of 
freedom and use the Hybrid Monte Carlo algorithm to explore 
thermodynamically important ${\rm Mn}$ spin configurations.
The critical temperature $T_c$ of the model is unambiguously
signalled in our finite-size simulations 
by pronounced peaks in fluctuations of both 
${\rm Mn}$ and band carrier total spins. 
The $T_c$'s we obtain are, over much of the model's parameter space, 
substantially smaller than those estimated using 
mean-field theory.  When mean-field theory fails, short-range
magnetic order and finite {\em local} carrier spin polarisation are present 
for temperatures substantially larger than $T_{c}$. 
For the simplest version of the model, which has a single
parabolic band with effective mass $m^*$, 
the dependence of $T_{c}$ on $m^*$ is sublinear at large
masses, in disagreement with the mean-field 
theory result $T_{c} \propto m^*$. \\

%
%
\end{abstract}


\section{Introduction}

Because their electronic properties are, at low temperatures,
extremely sensitive to external
magnetic fields, diluted magnetic semiconductors
have long attracted attention\cite{revDMS}. 
The discovery of ferromagnetism at comparatively high temperatures
($T_{c}>100{\rm K}$) in Mn-doped (III,V) semiconductors 
\cite{Ohn:96,Ohn:98} has intensified  
interest, partly because of the new technological pathways  
that might be opened if room temperature ferromagnetism
were achieved in a semiconductor with favorable materials
properties\cite{Ohn:SSCreview,Pri:98}. 
The search for materials in this class with higher 
ferromagnetic transition temperatures, and the effort to achieve a 
deeper understanding of all physical property trends, 
has stimulated a large body of experimental and theoretical research. 

To date, most theoretical work\cite{dietl_mft,Ramin,us_mft,sham_mft}
has been based on kinetic exchange 
models of the type used here and detailed below.  First principles
density-functional-theory electronic structure
calculations\cite{sanvito,butler}
are qualitatively in accord with these phenomenological models,
although more work may be necessary to reliably judge the accuracy 
of the local-spin-density approximation for these materials before
they can be used to guide model refinement.
The model's low-energy degrees of freedom are $S=5/2$ spins representing 
${\rm Mn}^{2+}$ ions and holes in the semiconductor valence band 
that are free to move through the 
system.  Ferromagnetism is most easily
discussed\cite{dietl_mft,Ramin,us_mft,sham_mft} 
in a Mn continuum version of the model, justified in part by
the small ratio of hole density to Mn ion density in experimental 
systems.  The constant Mn density model is
somewhat analogous to the jellium model for metals in that 
it artificially preserves translational invariance.  
In modelling these materials, the simplest plausible approximation 
is one that both represents local moments by a uniform density continuum
and ignores correlations between local moment orientations and 
the state of the valence-band-hole gas.  In the following we refer 
to this combination of idealizations as the mean-field-approximation.
In the mean-field-approximation\cite{dietl_mft,Ramin,us_mft,sham_mft},
the ground state is always ferromagnetic because the
exchange energy gained by polarizing is linear in the hole-spin density
whereas the band and hole-hole interaction energy cost is quadratic.
For the Mn concentrations at which the highest ferromagnetic transition
temperatures occur, it appears that mean-field-theory predictions for 
$T_c$, magnetic anisotropy, optical absorption, and other properties 
are at least qualitatively in accord with experiment,
although further experimental and theoretical work is needed to 
confidently establish where this approach is and is not reliable.

It is well known that mean-field theories tend to overestimate
the stability of ordered phases, because they fail to account 
for thermal and quantum fluctuations that can destroy long-range order. 
In particular, they predict critical temperatures which are systematically too 
high.  A step toward a more complete description of ferromagnetism in diluted
magnetic semiconductors was taken in Ref.~\cite{KLM:00}, where a theory of 
collective spin-wave excitations was developed for the continuum version 
of the kinetic-exchange model.
$T_c$ estimates \cite{SKLM:00} based on this theory of the
ferromagnet's spin stiffness establish that mean-field theory
becomes less reliable for stronger exchange coupling and for
flatter semiconductor valence bands; the higher 
the predicted $T_c$ the lower its reliability.
The approximate collective-fluctuation 
$T_c$ bounds proposed in Ref.~\cite{SKLM:00} are 
based on $T=0$ spin-wave energies of the continuum version of the kinetic 
exchange model.  The considerations of this paper demonstrate that 
collective magnetization fluctuations will control the temperature
at which long range magnetic order is lost  
in at least a part of the model-parameter range relevant to
ferromagnetic semiconductors.
In this work we report on Monte Carlo simulations,  
based on a Hybrid Monte Carlo algorithm \cite{HMC}, that 
fully account for thermal fluctuations of ${\rm Mn}$ spin orientations 
and shed further light on the finite-temperature properties of 
the kinetic-exchange model.  Some of the results of these calculations
were described briefly in Ref.~\cite{SKLM:00}.
In this paper we give a complete description of the calculations, 
including all important technical details. 
In addition, we also present results obtained using a 
realistic six-band $\vec k\cdot\vec p$
model for the valence bands of zincblende semiconductors,
allowing us to assess the importance for thermal fluctuations of 
band structure details.
A related Monte Carlo study by Sakai {\it et al.}\cite{sakai}
has the same motivation as ours, but 
examines a model of Ising localized spins coupled to band electrons.
This model has gapped collective excitations and qualitatively different 
thermodynamic properties.

Our simulations are directed toward the regime of ${\rm Mn}$ concentration
in ${\rm III}_{1-x}{\rm Mn}_x{\rm V}$ alloys where the highest
ferromagnetic transition temperatures have been observed,
$x \sim 0.05$ \cite{Ohn:98}. 
In this regime the heavily-doped semiconductors are 
strongly disordered three-dimensional metals with $k_F \ell \sim 3$.
(Here $k_F$ is a typical Fermi wavevector and $\ell$ is a mean-free
path estimated from the measured $T=0$ resistivity.)  The sources
of disorder in these alloys have not been fully cataloged, although
randomness in the Mn alloy positions and antisite defects 
(group V elements on group III sites) are certainly among 
them.  In our simulations we include only the former 
disorder source.  In the continuum version of the model this source 
is absent and the system is disorder-free.  We find 
that disorder has quantitative but not qualitative importance for
the model we study.   
Randomness in the Mn ion distribution will be increasingly important 
at lower Mn concentrations, particularly on the insulating side 
of the metal insulator transition which occurs for $x \sim 0.02$, 
and has an overriding importance in the models studied in Ref.~\cite{Bhatt} 
that are directed to the dilute Mn concentration limit. 

In the extreme dilute limit, all holes are bound to ${\rm Mn}^{2+}$ acceptors, 
forming neutral complexes. 
For small Mn concentrations, holes may hop from acceptor site to 
acceptor site, but it is still useful to regard them as belonging
to an impurity band.  On the metallic side of the metal insulator
transition, the concept of an impurity band is less useful
and the host semiconductor band structure has to be modelled realistically,
as in the present work.  
In order to correctly capture the physics of the dilute limit, 
we would need to include Coulombic interactions
between Mn ions and valence band spins in addition to 
exchange interactions.  If hole-hole interactions were treated in a 
mean-field-theory, including Coulomb scattering would add a 
self-consistently determined scalar potential to the band-electron
Hamiltonian.  There are several obstacles to including this 
scalar potential in our work at present.   The most fundamental
problem is current ignorance about the distribution of 
charged defects in these materials.  It is known that the 
hole density in typical samples is approximately three times
smaller than the Mn density and suspected that the compensation
is due mainly to the antisite defects present in low-temperature-growth 
compound semiconductors \cite{Omiya}. 
These defects will play as 
important a role as the Mn ions in determining the screened 
scalar potential.  There are also technical obstacles to 
including Coulomb interactions in our Monte Carlo calculations.
If they were explicitly included, we would have to 
self-consistently solve for the valence band electron screening 
charges for each Mn spin configuration, or adopt a Car-Parrinello \cite{CP}
type of scheme. In the present study we shall therefore ignore 
Coulomb interactions, partially to enable progress on a reasonable
level of technical difficulty warranted by the present status of 
experimental insight and  microscopic theoretical modelling.
This approximation dominates so far the interpretation of
experimental results on magnetic properties of Mn-doped III-V semiconductors,
and as well the theoretical discussion.

This paper is organized as follows: In section \ref{themodel} we discuss
the kinetic-exchange model of Mn-doped (III,V) semiconductors. In section 
\ref{HMCalg}
we summarize the main features of the Hybrid Monte Carlo algorithm that are
important for our purposes. Numerical results obtained by this method as well 
as a detailed discussion of important technical aspects are presented in 
section \ref{numres}.  Our data show very clearly the typical features of 
ferromagnetic transitions signalled unambiguously by susceptibility peaks. 
We discuss the critical temperature as a function of effective mass, carrier
density, and exchange-coupling strength.
We close with a summary in section V.


\section{The model}
\label{themodel}

We now describe the kinetic-exchange model we use for 
(III,Mn)V diluted magnetic semiconductor ferromagnets in detail.
In this material class, the highest critical temperature of 110K has been
obtained for ${\rm Ga}_{1-x}{\rm Mn}_x{\rm As}$ with 
$x\approx 0.05$ \cite{Ohn:98}.  The manganese ions are assumed to
be ${\rm Mn}^{2+}$ with a $3d^5$, $S=5/2$ configuration, and to be placed at 
random on group III element sites so that they are acceptors.
The concentration of holes is, however, allowed to differ from
the Mn concentration, since the system is strongly compensated by
anti-site defects, i.e. group V atoms on group III sites.  In practice
the hole concentration is often not accurately known; its
determination from Hall effect measurements is complicated by the
extraordinary Hall effect that occurs in all ferromagnetic metals
\cite{Omiya}.

Long range magnetic order occurs in this dilute system of magnetic moments 
because of an exchange interaction between the spins of the manganese ions and 
spins of the valence band holes.  
The minimal model to describe this carrier-induced ferromagnetism is
\begin{equation}
{\cal H}=\sum_{i}\frac{\vec p_{i}\,^{2}}{2m^{\ast}}
+\sum_{I}\int\,d^{3}r\,J(\vec r-\vec R_{I})\vec s(\vec r\, ) \cdot \vec S_{I}
\, .
\label{defmod1}
\end{equation} 
This Hamiltonian describes noninteracting carriers in a parabolic band 
characterized by an effective mass $m^{\ast}$, whose spin density
$\vec s(\vec r\, )$ is coupled to manganese ion spins $\vec S_{I}$
at locations $\vec R_{I}$ by an antiferromagnetic exchange coupling. 
To account for the finite extent of the manganese ions,
the exchange is modelled by a spatially extended coupling 
$J(\vec r\, )$ \cite{Bhatta:00}; we use a Gaussian centered around the ion 
position,
\begin{equation}
J(\vec r \,)=\frac{J_{pd}}{(2\pi a_{0}^{2})^{\frac{3}{2}}}
e^{-\frac{r^{2}}{2 a_{0}^{2}}}\,.
\label{excpl}
\end{equation}
Both the strength and range of this interaction are phenomenological 
parameters to be fixed by comparison with experiment or, ideally,
to be extracted from first principles electronic structure calculations.
The exchange-coupling range parameter $a_0$ in Eq.~(\ref{excpl}) is required 
in our calculations to keep exchange-coupling shifts in 
quasiparticle energies finite.  In the numerical studies to be presented below 
we usually choose $a_{0}=0.1{\rm nm}$, which appears to be a reasonable value 
for the d-ion radius in  ${{\rm Ga}_{1-x}{\rm Mn}_{x}{\rm As}}$, whose lattice 
constant is $\sim 0.60{\rm nm}$ \cite{Ohn:96,Ohn:98}.
However, we will see that properties of the system depend only very weakly on
$a_{0}$, as long as this quantity is not increased to large values.

The exchange parameter $J_{pd}>0$ characterizes the strength of the 
exchange interaction between carriers and manganese ions. Typical
experimental values for this quantity are of the order of $0.1 {\rm eVnm^{3}}$.
(The $pd$ subscript on the coupling constant is a reminder of its
presumed physical origin in the hybridization of valence-band 
p-orbitals and Mn-ion d-orbitals.  
Since the carrier spin density $\vec s(\vec r\, )$ has dimension 1/volume and 
$\vec S_I$ is dimensionless, $J_{pd}$ has the dimension of energy times 
volume.)

We emphasize that the interaction in Eq.~(\ref{defmod1}) is of the isotropic 
Heisenberg type.  In a parabolic band model, total energies 
are therefore invariant under global rotations of all $\rm Mn$ spins
and slow variations in orientation have a small energy cost.  Even 
with the realistic band models discussed below,
the energy cost of global Mn spin rotations is still
very small\cite{Ramin}.  Models with Ising-like interactions\cite{sakai}
{\em do not} capture this essential element of (III,Mn)V ferromagnet
physics, although they are much more easily studied.

An important refinement of the above model, Eq.~(\ref{defmod1}),  
is obtained by replacing the single parabolic band by 
a realistic model for the host semiconductor valence bands.
In the materials of interest these derive from atomic p-orbitals
and have a spin-orbit coupling strength that exceeds the 
Fermi energy.  
Importantly, eigenstates do not have definite angular momentum 
quantum numbers at finite wavevector; at the center of the
Brillouin zone, however, they have definite total angular momentum 
$J\in\{3/2,1/2\}$.
We use these states in a $\vec k\cdot\vec p$ scheme, choosing as 
basis states \cite{ChKo:99,Ramin}
\begin{eqnarray}
|1\rangle: & = & |\frac{3}{2},\frac{3}{2}\rangle=
|1,\uparrow\rangle\nonumber\\
|2\rangle: & = & |\frac{3}{2},-\frac{1}{2}\rangle= 
\frac{1}{\sqrt{3}}|-1,\uparrow\rangle
+\sqrt{\frac{2}{3}}|0,\downarrow\rangle\nonumber\\
|3\rangle: & = & |\frac{3}{2},\frac{1}{2}\rangle= 
\frac{1}{\sqrt{3}}|1,\downarrow\rangle
+\sqrt{\frac{2}{3}}|0,\uparrow\rangle\nonumber\\
|4\rangle: & = & |\frac{3}{2},-\frac{3}{2}\rangle=
|-1,\downarrow\rangle\nonumber\\
|5\rangle: & = & |\frac{1}{2},\frac{1}{2}\rangle= 
-\frac{1}{\sqrt{3}}|0,\uparrow\rangle
+\sqrt{\frac{2}{3}}|1,\downarrow\rangle\nonumber\\
|6\rangle: & = & |\frac{1}{2},-\frac{1}{2}\rangle= 
\frac{1}{\sqrt{3}}|0,\downarrow\rangle
-\sqrt{\frac{2}{3}}|-1,\uparrow\rangle.
\end{eqnarray}
Near the Brillouin-zone center, the band structure can be parameterized by a
small number of symmetry-adapted parameters of the   
Kohn-Luttinger Hamiltonian \cite{LuKo:55,ChKo:99,Ramin}). In the above
basis this effective kinetic-energy operator reads 
\begin{equation}
\hspace*{0cm} {\cal H}_{KL} 
= \left(\begin{array}{cccccc} {\cal H}_{hh} & -c & -b &
0 & \frac{b}{\sqrt{2}} & c\sqrt{2}\\ -c^* & {\cal H}_{lh}
& 0 & b & -\frac{b^*\sqrt{3}}{\sqrt{2}} & -d\\ -b^* & 0 &
{\cal H}_{lh} & -c &   d & -\frac{b\sqrt{3}}{\sqrt{2}} \\
0 & b^* & -c^* & {\cal H}_{hh} &  -c^*\sqrt{2} &
\frac{b^*}{\sqrt{2}}\\ \frac{b^*}{\sqrt{2}} &
-\frac{b\sqrt{3}}{\sqrt{2}} & d & -c\sqrt{2} & {\cal H}_{so} & 0\\
c^*\sqrt{2} & -d & -\frac{b^*\sqrt{3}}{\sqrt{2}} & \frac{b}{\sqrt{2}} & 0 &
{\cal H}_{so}\\
\end{array}\right)
\label{KL}
\end{equation}
where the entries of this 6$\times$6 matrix for each wavevector $\vec k$ are
given by 
\begin{eqnarray}
{\cal H}_{hh} &=& \frac{\hbar^2}{2m}\left[(\gamma_1 + \gamma_2)(k_x^2+k_y^2) +
(\gamma_1 - 2\gamma_2)k_z^2\right]\nonumber \\ {\cal H}_{lh} &=&
\frac{\hbar^2}{2m}\left[(\gamma_1 - \gamma_2)(k_x^2+k_y^2) + (\gamma_1 +
2\gamma_2)k_z^2\right] \nonumber \\ {\cal H}_{so} &=&
\frac{\hbar^2}{2m}\gamma_1(k_x^2+k_y^2+k_z^2) + \Delta_{so} \nonumber \\ b &=&
\frac{\sqrt{3}\hbar^2}{m} \gamma_3 k_z (k_x - i k_y) \nonumber \\ c &=&
\frac{\sqrt{3}\hbar^2}{2m}\big[\gamma_2(k_x^2 - k_y^2) - 2i\gamma_3 k_x
k_y\big] \nonumber \\ d &=&
-\frac{\sqrt{2}\hbar^2}{2m}\gamma_2\big[2k_z^2-(k_x^2 + k_y^2 )\big]\; .
\label{lutpar}
\end{eqnarray}
The quantities $\gamma_{1}$, $\gamma_{2}$, $\gamma_{3}$ are called Luttinger 
parameters, $m$ is the bare electron mass, and $\Delta_{so}$ is the spin-orbit 
coupling which splits the six states at the valence band edge into a doublet 
and a quartet.  These band parameters are accurately known for a large number 
of (III,V) compound semiconductors \cite{meyer}.
For all numerical calculations we use
$(\gamma_{1},\gamma_{2},\gamma_{3})=(6.85,2.1,2.9)$ and 
$\Delta_{so}=0.34{\rm eV}$, the band parameter set appropriate for GaAs.

In the kinetic-exchange model with realistic band structure, the 
Hamiltonian~(\ref{KL}) replaces the kinetic part of Eq.~(\ref{defmod1}).
The Hilbert space dimension of the one-particle problem is, thus, enlarged by 
a factor of three.
The Kohn-Luttinger Hamiltonian has already been adopted in some mean-field 
calculations for the continuum kinetic exchange 
model\cite{dietl_mft,Ramin}.
This realistic description is essential if important 
magnetic parameters connected with magnetocrystalline anisotropy and 
anisotropic magnetoresistance are to be addressed.

The material ${\rm Ga}_{1-x}{\rm Mn}_{x}{\rm As}$ has a zincblende structure, 
in which the cations and anions each form an fcc lattice.  In the alloy
manganese ions are located on randomly chosen cation sites,
forming a disordered system of local 
moments.  In an approximation consistent with our band Hamiltonians we ignore 
the underlying lattice and place the ions positions completely at random 
within periodically continued cubic simulation cells.
In the Monte Carlo calculations described below, we average over different 
disorder realisations.  However, we find that the results of different 
realizations typically lie within a range of a few percent.

In our simulations we treat the $S=5/2$ Mn spins as classical degrees
of freedom.  This approximation is not severe even in a mean-field theory 
where it alters the critical temperature by a factor of $1 + 1/S =1.4$.
It is even less important in the common circumstance where the most important 
spin fluctuations are collective.  
This classical approximation is, therefore, a less serious source of 
uncertainty in our representation of (III,Mn)V ferromagnet
thermodynamics than the modelling issues discussed above; in our 
judgement the substantially greater complexity of quantum 
calculations is not warranted at present.
Thus, the state of each ion spin $\vec S_{I}$ is specified by two polar angles 
$\vartheta_{I}$, $\varphi_{I}$. 
In interpreting results, however, it is important to realize that 
some low-temperature power laws will be altered by quantum freeze out.

We are interested in thermal expectation values
of the form 
\begin{equation}
\bar f =
\frac{1}{\cal Z}
\int_{0}^{2\pi}d\varphi\int_{0}^{\pi}d\vartheta
\sin\vartheta\,
{\rm Tr}\left\{\hat f(\vartheta,\varphi)e^{-\beta{\cal H}}\right\}\,,
\label{exval1}
\end{equation}
where $\beta$ is the inverse temperature, $\cal Z$ the partition function,
and $\vartheta$, $\varphi$ are
shorthand notations for the whole set of classical spin coordinates.
The quantity $\hat f(\vartheta,\varphi)$ is a function of the ion spin
angles and an operator with respect to the quantum mechanical carrier degrees
of freedom over which the trace is performed.
In practice we replace the fermion trace by a ground state 
expectation value, since the temperatures of interest will always be much
smaller than the Fermi energy.  For typical carrier 
densities $n$ of order $0.1{\rm nm}^{-3}$,
the Fermi temperature for the carriers is $ \gtrsim 1000{\rm K}$,
compared to ferromagnetic critical temperatures $ \sim 100{\rm K}$.
Therefore, the average over the carrier system may be approximated
by an average over its $T=0$ ground state. Thus, 
\begin{equation}
\bar f = \frac{1}{\cal Z}
\int_{0}^{2\pi}d\varphi\int_{0}^{\pi}d\vartheta
\sin\vartheta\,
\langle 0|\hat f(\vartheta,\varphi)|0\rangle
e^{-\beta\langle 0|{\cal H}|0\rangle}\,,
\label{exval2}
\end{equation}
where $|0\rangle$ denotes the ground state of non-interacting 
fermions with the appropriate band Hamiltonian and a Zeeman-coupling term 
$h$ whose effective magnetic field $B_{\rm eff}$ is due to exchange 
interactions with the localized spins,
\begin{eqnarray}
h &=& - \int d^{3}r\, \vec s(\vec r \, ) \cdot \vec B_{\rm eff}(\vec r \, ) 
\nonumber \\
\vec B_{\rm eff}(\vec r\, ) 
&=& - \sum_{I} J(\vec r - {\vec R}_{I}) S  {\hat \Omega}_I
\label{eq:beff}
\end{eqnarray}
where $\hat\Omega_I 
=(\sin(\theta_I)\cos(\phi_I),\sin(\theta_I)\sin(\phi_I),\cos(\theta_I))$
is the direction of the classical spin at ${\vec R}_I$.
In the following we denote thermal expectation values of quantities defined 
in terms of classical spin orientation variables 
by $\langle\cdot\rangle$ and quantum mechanical expectation values within 
the carriers ground state by $\langle 0|\cdot|0\rangle$. 
The latter quantities will be averaged also thermally according to 
Eq.~(\ref{exval2}).


\section{The Hybrid Monte Carlo Algorithm}
\label{HMCalg}

A standard way to evaluate expectation values of the form Eq.~(\ref{exval2}) 
is to use classical Monte Carlo algorithms which perform a random walk in the 
phase space of the classical variables ($\vartheta$,$\varphi$). 
The probabilities governing this
Monte Carlo dynamics are specified by the dependence of many-fermion
energy on the localized-spin configuration.
The many-fermion ground state is a Slater determinant whose single-particle 
orbitals are the lowest energy eigenstates of a single-band or multi-band 
Hamiltonian of the form (\ref{defmod1}). For the case of a parabolic
band, the matrix elements of the corresponding one-particle Hamiltonian 
in a plane-wave basis read
\begin{eqnarray}
& & \langle\vec k^{\prime} \sigma^{\prime}|{\cal H}|\vec k \sigma\rangle 
=\frac{\hbar^{2}k^{2}}{2m^{\ast}}\delta_{\vec k,\vec k^{\prime}}
\delta_{\sigma,\sigma^{\prime}} \nonumber\\
& & \qquad\qquad\qquad
+\frac{S}{2L^{3}}\sum_{I}J(\vec k-\vec k^{\prime})
e^{i(\vec k-\vec k^{\prime})\vec R_{I}}\hat\Omega_{I} \cdot
\vec\tau_{\sigma,\sigma^{\prime}}\,,
\label{singpart}
\end{eqnarray}
where $\vec k$ and $\sigma$ denote wavevector and spin indices, respectively,
$J(\vec k\, )$ is the Fourier transform of $J(\vec r\, )$, and $L$ the edge 
length of the simulation cube. Periodic boundary conditions restrict the 
admissible values of wavevector components to integer multiples 
of $2\pi/L$. In Eq.~(\ref{singpart}) $\vec\tau$ is the vector of Pauli
spin matrices. 

Since the many-particle groundstate of the carrier system has to be
computed at each Monte Carlo step, the computational effort required for
the present calculations is much larger 
than in simple classical spin models.  In the usual Metropolis algorithm,
a single spin orientation is altered at each step.  If this algorithm
were employed here the time required to diagonalize the single-particle
Hamiltonian each time would severely limit the efficiency of the algorithm. 
We therefore use the Hybrid Monte Carlo algorithm, which was introduced in 
the mid 1980's in the context of lattice field theories \cite{HMC}.
In this method {\em all} classical variables are altered in one Monte Carlo 
step.
This drastically reduces the number of matrix diagonalisations required to 
explore statistically important magnetic configurations.
The Hybrid algorithm is a powerful method for Monte Carlo simulations in 
systems containing coupled classical and quantum mechanical degrees of freedom.
Very recently, a variant of this algorithm has been applied to a system
of classical degrees of freedom coupled to non-interacting lattice fermions
\cite{AFGLM:00}, a problem similar to the one studied here. 

The Hybrid Monte Carlo algorithm determines average values defined  
in terms of a probability distribution $P$ of the form
\begin{equation}
P(\phi)\propto e^{-{\cal S}(\phi)} \, .
\label{candis}
\end{equation}
The ``action'' ${\cal S}$ depends on a set of classical variables summarized 
by the symbol $\phi$. 
The basic trick of the algorithm is to introduce a ``fake dynamics'' for the
classical variables which is governed by the ``fake Hamiltonian'' 
\begin{equation}
H^{\prime}=\frac{1}{2}\pi^{2}+{\cal S}^{\prime}(\phi) 
\end{equation} 
with ``fake momenta'' $\pi$
and an action ${\cal S}'$ which is not necessarily identical to 
${\cal S}$.

In this algorithm one Monte Carlo step is performed in the following way: 
(i) Choose a value for each fake momentum from its Gaussian distribution.
(ii) Let the system evolve in Monte Carlo time $\tau$ according to the 
Hamilton equations of motion,
\begin{equation}
\partial_{\tau}\phi=\frac{\partial H^{\prime}}{\partial \pi}=\pi\quad,\quad
\partial_{\tau}\pi=-\frac{\partial H^{\prime}}{\partial \phi}
=-\frac{\partial {\cal S}^{\prime}}{\partial \phi}\, ,
\label{fakedyn}
\end{equation}
which leads to a new configuration of fields and fake momenta 
$(\phi^{\prime},\pi^{\prime})$. 
(iii) Accept this new configuration with a probability
\begin{equation}
\min\{1,\exp[H(\phi,\pi)-H(\phi^{\prime},\pi^{\prime})]\}\,,
\label{accprob}
\end{equation}
where $H$ is the ``true Hamiltonian'',
\begin{equation}
H=\frac{1}{2}\pi^{2}+{\cal S}(\phi)\,.
\end{equation} 
This acceptance condition is reminiscent of the Metropolis algorithm. 
We note that the acceptance probability is strictly unity for 
${\cal S}={\cal S}^{\prime}$, 
and that the acceptance rate in this Monte Carlo procedure reflects the 
difference between the two actions ${\cal S}$ and ${\cal S}^{\prime}$.

Thermodynamic averages are calculated using 
\begin{equation}
\bar f=\frac{1}{N}\sum_{n=1}^{N}\langle 0_{n}|
\hat f(\vartheta_{n},\varphi_{n})|0_{n}\rangle\,,
\end{equation}
where $n$ labels the accepted configurations.
As shown in Ref.~\cite{HMC} this Hybrid Monte Carlo algorithm generates a 
Markov chain which converges to the distribution (\ref{candis}).

Our calculations adopt this scheme for the set of classical spin angles
and the action
\begin{equation}
{\cal S}=\beta\langle 0|{\cal H}|0\rangle-\sum_{I}\ln|\sin\vartheta_I|\, .
\label{action}
\end{equation}
The first term is the energy of the carrier system for a given classical spin 
configuration and the logarithms arise from the Jacobi determinant in
Eq.~(\ref{exval2}). Note that this part is not multiplied by the
inverse temperature $\beta$.

As pointed out above, the action ${\cal S}'$ need not to be the same as 
${\cal S}$. In our calculation, we use this freedom to reduce the computational
effort.  Our fake action $S'$ is defined as in Eq.~(\ref{action}) but
for an itinerant-carrier state $|0\rangle_{i}$ that is evaluated in the
beginning of each Monte Carlo step and taken to be fixed in the susequent
integration of the fake dynamics (\ref{fakedyn}).
In addition, we replace $J(\vec r\,)$ by $J_{pd} \delta(\vec r\,)$, i.e., 
we use
the $a_0\to 0$ limit of the exchange coupling for the fake dynamics.
The force term in the effective dynamics is then readily evaluated from
\begin{equation}
\frac{\partial 
\left(_{i}\langle0|{\cal H}|0 \rangle_{i}\right)}{\partial \hat \Omega_I}
= JS\left(_{i}\langle 0|\vec s(\vec R_{I})|0\rangle_{i}\right) \, .
\label{efffield}
\end{equation}
In the fake dynamics the localized spins move under the influence
of the fixed effective magnetic fields generated by exchange interactions
with the fixed electronic spins of $|0 \rangle_{i}$.
With this approach, only one matrix diagonalization is required to update
all Mn spin orientations.
However, since the acceptance probability (\ref{accprob}) is determined by the
action ${\cal S}$, the sampling is still performed with respect to the 
original distribution (\ref{candis}).
The choice of ${\cal S}'$ being different from ${\cal S}$ is, therefore, not 
an approximation as 
long as the Hamiltonian fake dynamics is reversible, a feature which can be 
incorporated in the numerical integration using
the leapfrog algorithm \cite{HMC}.

The action (\ref{action}) becomes singular if one of the polar angles 
$\vartheta$ is equal to an integer multiple of $\pi$. 
This may lead, in principle, to difficulties in the numerical implementation 
of our method. 
But since this singularity is only logarithmic, and since in the fake 
dynamics (\ref{fakedyn}) the $\vartheta$'s are repelled from the singular 
points, we have never encountered any practical difficulty due to these 
coordinate singularities.


\section{Numerical Results}
\label{numres}

In this section we present our numerical results.  We concentrate on the spin 
polarizations of the manganese ions and the carriers as a function of 
temperature and address the ferromagnetic transition. 
We start with the case of parabolic bands described by the 
Hamiltonian (\ref{defmod1}).

\subsection{Parabolic bands}

In the following we present results in 
dimensionful units for parameters in the range of interest for
(Ga,Mn)As.  It is hoped that ferromagnetism
will be realized in other materials described by the same model
but with different parameter values.  In this connection,
it is useful to observe that 
band and exchange interaction terms in our model Hamiltonian
both scale simply with changes in length scale and model 
parameters.  Results for the parabolic band model have a 
non-trivial dependence only on the ratio of the density $n$ band carriers 
and the concentration of Mn ions $N$, and on the ratio of 
exchange interaction and band energy scales, $\Delta/\epsilon_F$.  
Here $\Delta = J_{pd} N S$ is the continuum model
mean-field-theory band exchange splitting, and 
$\epsilon_F/ k_B  = T_F \approx 4230 {\rm K} (m/m^*) (n[{\rm nm}^{-3}])^{2/3}$
is the band Fermi energy in the paramagnetic state.
There is also a very weak dependence on the dimensionless range of the 
exchange interaction $a_0 N^{1/3}$, that does not play an important 
role and will usually be ignored.

In the following subsections we first show some typical numerical results 
for magnetization averages and fluctuations and explain how we 
extract $T_c$ from them.  We then proceed with some 
technical considerations that enter into these calculations before 
summarizing the numerical results we have obtainted to date.

\subsubsection{Magnetisation curves and fluctuations}

Figure~\ref{fig1} shows typical magnetisation data as a function of the
temperature. These results were obtained for a maganese ion density of
$N=1.0{\rm nm^{-3}}$, a carrier density $n=0.1{\rm nm^{-3}}$ in a cubic 
simulation volume of $V=540{\rm nm^{3}}$, i.~e. the system contains 
540 maganese
ions and 54 carriers. The effective band mass is half the bare electron mass,
and the exchange parameter is $J_{pd}=0.15{\rm eVnm^{3}}$. 
The main panel shows the average polarisation of the manganese spins,
\begin{equation}
M=\frac{1}{NV}\langle|\vec S_{tot}|\rangle\,,
\end{equation}
i.~e. the thermally averaged modulus of the total ion spin, along with
the carrier magnetisation,
\begin{equation}
m=\frac{1}{nV}\langle|\langle 0|\vec s_{tot}|0\rangle|\rangle\,,
\label{carrmag}
\end{equation}
which is the ensemble average of the modulus of the total ground-state carrier 
spin. Both quantities are divided by the number of 
particles and are close to their maximum values at low temperatures.
At higher temperatures they show the expected transition to a paramagnetic 
phase. The critical temperature of this ferromagnetic transition 
is most readily estimated from numerical results for the magnetisation
fluctuations:
\begin{eqnarray}
g_{Mn} & = & \frac{1}{NV}\left(\langle|\vec S_{tot}|^{2}\rangle
-\langle|\vec S_{tot}|\rangle^{2}\right)\,,
\label{fluc1}\\
g_{p} & = & \frac{1}{nV}\left(
\langle|\langle 0|\vec s_{tot}|0\rangle|^{2}\rangle
-\langle|\langle 0|\vec s_{tot}|0\rangle|\rangle^{2}\right)\,.
\label{fluc2}
\end{eqnarray}
These two fluctuations per particle are plotted in the insets of 
Fig.~\ref{fig1}. 
They both show a pronounced peak at a temperature $T \sim 100 {\rm K}$,
defining the finite-system transition temperature for these model 
parameter values.  
In fact, in a region around this transition both datasets differ just by 
a factor of approximately 25, which is the square of the ratio of the two 
spin lengths entering the expressions (\ref{fluc1}) and (\ref{fluc2}), 
respectively.
This observation shows explicitly that
the correlation length both for the manganese ions and in the carrier 
system is the same near the transition, namely given essentially by the
system size.

In conclusion, our Monte Carlo approach clearly reproduces the expected 
ferromagnetic transition. The transition temperature 
$T_c$ can be determined unambiguously and consistently 
from the positions of very pronounced peaks in total magnetisation 
fluctuations of both the Mn ions and the carriers.

\subsubsection{Technical Considerations}

\begin{figure}
\centerline{\includegraphics[width=8cm]{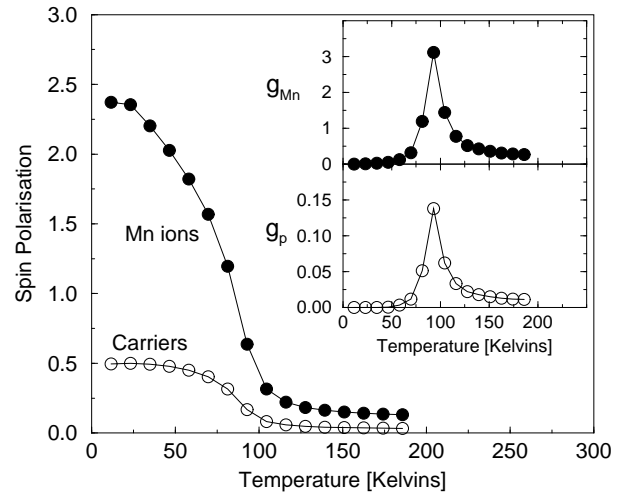}}
\caption{Magnetization curves for manganese ions and carriers.
The upper and lower inset show the magnetic fluctuations for the manganese
ions and the carriers, respectively. Both differ by a factor of approximately
25 reflecting the square of the ratio of spin lengths.
The density of manganese
ions is $N=1.0{\rm nm^{-3}}$, the carrier density is 
$n=0.1{\rm nm^{-3}}$ in a cubic volume of $V=540{\rm nm^{3}}$.
The band mass is half the bare electron mass with an exchange parameter of
$J_{pd}=0.15{\rm eVnm^{3}}$.
\label{fig1}}
\end{figure}

\begin{figure}
\centerline{\includegraphics[width=8cm]{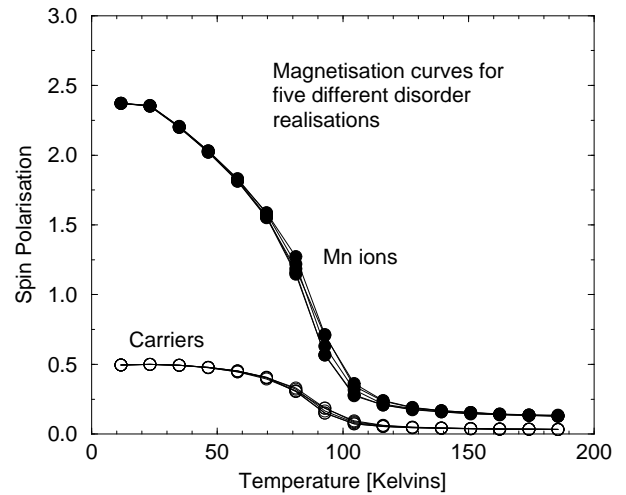}}
\caption{Magnetisation curves for five different realisations of manganese
positions underlying the averaged data of Fig.~\protect{\ref{fig1}}.
\label{fig2}}
\end{figure}

(i) {\bf Monte Carlo parameters and disorder realisations.} The data of
Fig.~\ref{fig1} was obtained by averaging over five different
realisations of the manganese-ions. For each
Monte Carlo run the system was thermalized within the first
$1000$ accepted steps, and the subsequent measurement phase included
$10000$ accepted steps. 
The Hamilton equations of motions (\ref{fakedyn}) are integrated over an 
interval of length $1$ in each Monte Carlo step.
These simulation parameters were used for all results reported in this paper. 
After the first $1000$ steps the magnetisations for the five different systems 
differ only by a few percent.
This indicates that the thermalization phase was long enough. In fact,
for not too large density ratios $n/N$, the final results for the
different disorder realisations differ only very weakly from each other.
This is illustrated in Fig.~\ref{fig2}, where the magnetisation curves
underlying the averaged results of Fig.~\ref{fig1} are plotted.
Those five datasets are hardly distinguishable from each other. 
 
Finally we mention that possible errors bars due to the statistical
uncertainty in the magnetisation data
for individual disorder realisations are by far smaller than
the symbols used in the diagrams. This follows from the analysis
of the Monte Carlo magnetisation data as a function of Monte Carlo time
which shows that the data is very well-converged for the algorithm parameters
given above, an observation which is of course consistent with 
(and actually a necessary condition for) the 
very smooth form of of our data plots as a function of temperature.

\begin{figure}
\centerline{\includegraphics[width=8cm]{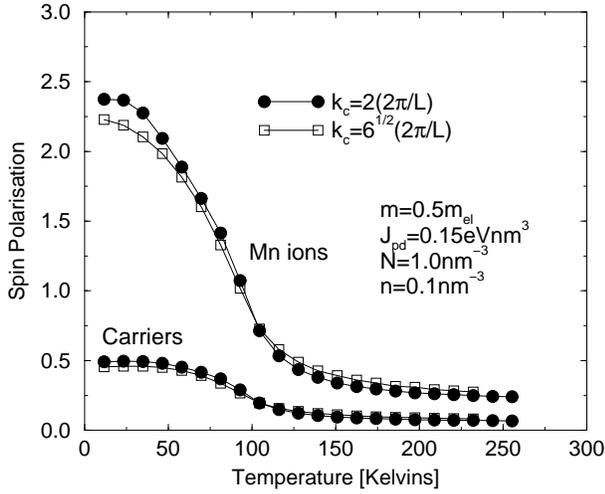}}
\caption{Magnetisation data for the same system parameters as in
Fig.~\protect{\ref{fig1}} in a cubic volume of $V=140{\rm nm^{3}}$ for
two different values of the wavevector cutoff. In a large area around
the ferromagnetic transition both datasets are very close two each other.
\label{fig3}}
\end{figure}

(ii) {\bf Wavevector cutoff.} Periodic boundary conditions restrict 
wavevector components for the carriers to integer multiples of $2\pi/L$, 
where $L=V^{1/3}$ is the edge length of the simulation cube. 
In our numerical calculations we include wavevectors $\vec k$ up to a certain
cutoff $k_{c}$. For the single-parabolic band model
the wavevector cutoff corresponds simply to a kinetic-energy cutoff.
The wavevector cutoff has to be chosen carefully, since our results are 
most cleanly interpreted when their cutoff dependence has been saturated.
On the other hand the dimension of the single-particle
Hamiltonian matrices, whose construction limits our numerical
calculations, grows with the third power of the cutoff.
Fortunately, our data converge comparatively quickly
with the cutoff parameter.
Figure~\ref{fig3} shows magnetization curves for the same system parameters 
as in Fig.~\ref{fig1} (but in a smaller simulation cell of volume 
$V=140{\rm nm^3}$) for two different cutoff parameters. 
In the first case, $k_{c}=2(2\pi/L)$
and the single-particle Hilbert 
space for the 14 carriers has dimension 66 (including the spin degree of 
freedom), while in the second case,
$k_{c}=\sqrt{6}(2\pi/L)$ and the dimension increases to 162.
The datasets shown in Fig.~\ref{fig3} are very close to each other over a large
temperature range around the ferromagnetic transition. This demonstrates that 
the smaller cutoff is already sufficient (nevertheless we used the larger one
for Fig.~\ref{fig1}).  
In the simulations of larger systems with more carriers the cutoff
parameter is appropriately adjusted to a value of $k_{c}=\sqrt{6}(2\pi/L)$.

\begin{figure}
\centerline{\includegraphics[width=8cm]{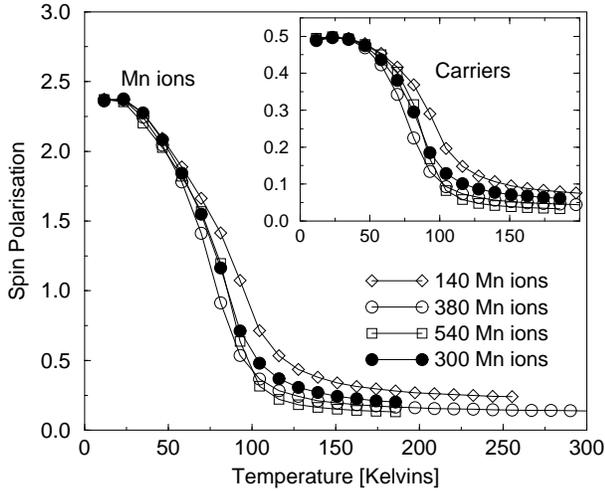}}
\caption{Magnetisation data for the same densities and
band mass and exchange parameter as in
Fig.~\protect{\ref{fig1}} for different system sizes. The first three
systems with 14, 38, and 54 carriers correspond to ``closed shell''
configurations in the paramagnetic carrier state, while the last system
(30 carriers) lies in between.
\label{fig4}}
\end{figure}

(iii) {\bf Finite-size effects.} 
The finite size of the simulation cell implies that the 
set of fermion energies obtained for a particular Mn spin
orientation configuration is discrete.
The gaps between adjacent energy eigenvalues depends on the size of the 
simulation cube and may lead to systematic $\vec k$-space shell effects 
in the model's finite-size dependence.  These are less important 
at temperatures of interest, since the typical orientation
configuration is complex and produces an exchange field
that removes all degeneracies and smears the $\vec k$-space shells.
Shell effects are therefore more important for weaker exchange coupling,
limiting our ability to do reliable calculations in the weak-coupling limit.
As we discuss later, however, we believe that mean-field theory
tends to be reliable in this limit, reducing the motivation 
for Monte Carlo studies.
Nevertheless, finite-size effects can be observed at all coupling 
strengths.  Their character is easily understood by recognizing
that Mn spins are coupled by the polarization they produce
in the electron system.  When a $\vec k$-space shell is partially
filled, its degeneracy is lifted by exchange copuling with 
a Mn spin-configuration.  Occupying the lower energy states in 
the split multiplet produces a larger electron spin polarization 
than when the multiplet is full.  Of course these effects 
become less important at larger system sizes when the typical
level shift is much larger than the typical level spacing.
The ferromagnetic transition temperature has systematic minima 
for the closed-shell `magic' electron numbers at which 
most of our simulations are performed.
This effect becomes smaller with increasing system size. 
Figure~\ref{fig4} shows magnetisation curves for the same densities
and system parameters as in Fig.~\ref{fig1} for systems with 14, 38, 54, and 30
carriers. The first three carrier numbers correspond to ``closed shell''
configurations in the paramagnetic phase while the last one lies in between.
The comparison of the data for the three largest carrier numbers shows
that the finite-size effect has already become weak for carrier numbers 
$nV\gtrsim 30$, while the magnetisation
curve for the system of 14 carriers shows slightly larger 
values at given temperature. However, the finite-size critical
temperature obtained in the case lies only by about $15{\rm K}$ above
the values for the other three system sizes.

\begin{figure}
\centerline{\includegraphics[width=8cm]{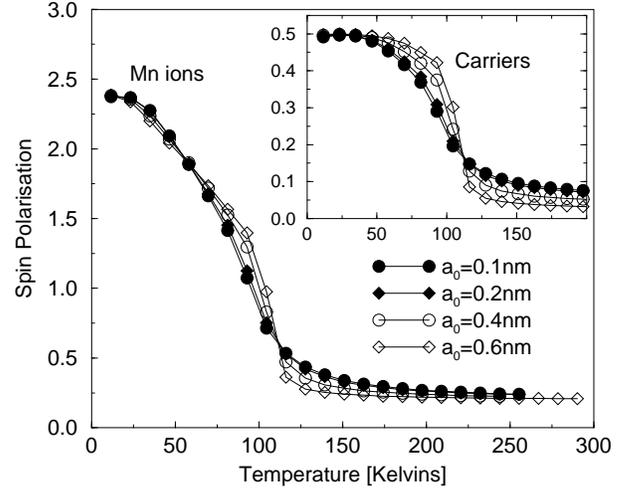}}
\caption{Magnetisation data for the same system parameters as in
Fig.~\protect{\ref{fig1}} in a cubic volume of $V=140{\rm nm^{3}}$ for
different values of the regularisation parameter $a_{0}$. At not too large
values for $a_{0}$ the dependence on this quantity is extremely weak.
\label{fig5}}
\end{figure}

(iv) {\bf The regularisation parameter $a_{0}$.} As mentioned above
we have chosen the regularisation parameter in the exchange coupling 
(\ref{excpl}) to be $a_{0}=0.1{\rm nm}$. 
This seems to be a reasonable value compared with the lattice constant of 
GaAs. Figure~\ref{fig5} shows the magnetisation data for the same system 
parameters as in Fig.~\ref{fig1} and a simulation volume of $V=140{\rm nm^3}$ 
for different values of $a_{0}$. 
The data for $a_{0}=0.1{\rm nm}$ and $a_{0}=0.2{\rm nm}$ are almost identical.
For larger values of $a_0$ a slight departure sets in. 
As a consequence, our results are practically independent of $a_0$ over the
(physically motivated) range of considered values.

In summary, we have demonstrated that our Monte Carlo results are,
in a large vicinity of $T_{c}$, 
stable with respect to the influence of the wavevector cutoff and the
finite system size. Moreover, the effect of different disorder realisations
and the precise value of the regularisation parameter of the
exchange coupling is found to be very weak.

\subsubsection{Results for $T_{c}$}

We now turn to the transition temperature $T_{c}$ for the parabolic band
model (\ref{defmod1}). Within mean field theory 
\cite{dietl_mft,Ramin,us_mft} this quantity is given by
\begin{equation}
T_c^{MF} = \frac{\chi_{P}}
{(g^{\ast} \mu_B/2)^2} \frac{S(S+1)NJ_{pd}^{2}}{12}\,,
\label{tcmf}
\end{equation}
where $S=5/2$ is the spin length of the manganese ions, 
and $g^{\ast}$ is the g-factor of the carriers. Their Pauli susceptibility
$\chi_{P}$ reads
\begin{equation}
\chi_{P}=\frac{3}{2}\frac{n}{\varepsilon_{F}}(g^{\ast}\mu_{B}/2)^2
\end{equation}
and is via the Fermi energy $\varepsilon_{F}$ proportional to the effective
mass $m^*$.   
Our objective here is not to make a quantitative prediction of
the critical temperature in particular ferromagnetic semiconductor 
systems.  In our view, uncertainly in model parameters and 
the possible importance of elaborations to the current model that  
account for example for direct Mn ion interactions, make such an effort 
premature.   By doing a numerically exact calculation for 
a model that captures much of the physics, however, we hope to
shed light on the range of validity and the sense and 
magnitude of likely corrections to mean-field-theory estimates.

The above expression for $T_{c}$ can be obtained by averaging the ion-spin and
carrier polarization over space.  The effective field which each manganese 
spin experiences due to a finite carrier polarization is constant in space and
the carrier bands are in turn spin split by $\Delta=J_{pd}NS$ by a spatially
homogeneous effective magnetic field.  
(The limit in which mean-field theory is exact can be achieved in our model 
by letting $a_0 \rightarrow \infty$ in Eq.~(\ref{excpl}).)
In this very simplified approach, spatial fluctuations and correlations 
between carriers and manganese spins are neglected.
As a result, mean-field theory predicts $T_{c}$ to be quadratic 
in the exchange parameter $J_{pd}$ and linear in the effective band mass $m^*$.
This mean-field theory has been used to predict values for $T_c$ 
in various semiconductor host materials\cite{dietl_mft}.
It is, however, expected \cite{KLM:00,SKLM:00} that due to the neglect of 
correlations, Weiss mean-field theory can substantially overestimate the 
critical temperature.

We therefore investigate the critical temperature $T_{c}$ with the help of
our Monte Carlo scheme. 
As seen in Fig.~\ref{fig1} this quantity can be read off unambiguously from 
the magnetic fluctuations of both the Mn ions and the carriers.
To limit computational expense we concentrate on systems with fourteen 
carriers.  
From Fig.~\ref{fig4} we conclude that finite-size effects will lead to values 
of $T_c$ which are only slightly too large (less than $20 {\rm K}$).
Furthermore, the {\em qualitative} behavior of the $T_{c}$ data as a 
function of system parameters such as the effective mass $m^*$ will not be 
affected by the finite simulation-cell size. 

\begin{figure}
\centerline{\includegraphics[width=8cm]{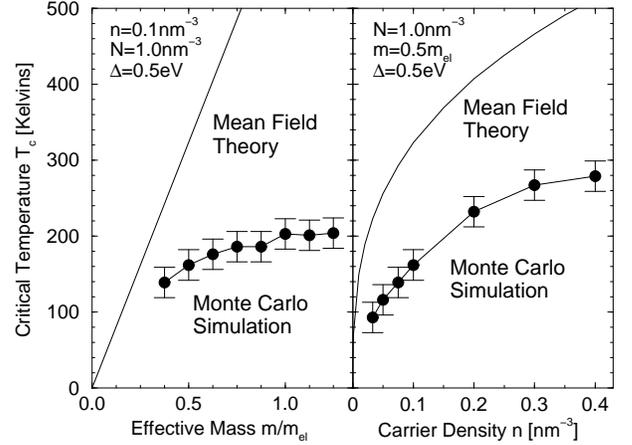}}
\caption{The critical temperature $T_{c}$ as a function of the carrier mass
(left panel) and the carrier density (right panel).
The exchange parameter is in both cases $J_{pd}=0.20{\rm eVnm^{3}}$
leading to a zero temperature mean-field spin splitting
of $\Delta=J_{pd}NS=0.5{\rm eV}$.
The results of the Monte Carlo runs are compared with the mean-field
predictions.
\label{fig6}}
\end{figure}

In Fig.~\ref{fig6} we show results for a manganese densities 
$N=1.0{\rm nm^{-3}}$ and a mean-field band splitting 
$\Delta=J_{pd}NS=0.5{\rm eV}$. 
The left panel shows the dependence of the critical temperature on the 
carrier effective mass. 
This dependence is very important for the search for diluted magnetic 
semiconductor systems with $T_{c}$'s larger than room temperature. 
In the mean-field approximation, $T_{c}$ grows linear with increasing mass. 
The Monte Carlo results clearly deviate from this prediction suggesting a 
saturation of $T_{c}$ at carrier masses close to the bare electron mass. 
For even higher masses, we expect the electrons to behave more 
classically and localize around individual Mn spins.  The cost 
in electronic energy of changing relative orientations of nearby
Mn spins will get smaller causing $T_c$ to decline and 
eventually causing ferromagnetism to disappear.  This physics
also occurs in a continuum model where the spin stiffness
declines as $1/m^*$ for large band masses\cite{KLM:00,SKLM:00},
although there are differences in detail.

To discuss the critical temperature as a function of the exchange coupling 
parameter $J_{pd}$ we observe that the Hamiltonian (\ref{defmod1})
satisfies the scaling relation
\begin{equation}
\beta{\cal H}\left(m^{\ast},J_{pd}\right)
=\frac{\beta}{q}{\cal H}\left(\frac{m^{\ast}}{q},q J_{pd}\right)
\label{scalerel}
\end{equation}
with $q>0$.
Therefore the saturation of $T_{c}$ as a function of the effective mass at 
fixed $J_{pd}$ corresponds to a linear dependence of $T_c$ on $J_{pd}$ at 
fixed $m^{\ast}$. This contrasts with the mean-field prediction
$T_c^{MF} \propto J_{pd}^2$.

In the right panel of Fig.~\ref{fig6} we show $T_{c}$ as a function of the 
carrier density. 
Here the Monte Carlo approach also clearly yields a lower critical
temperatures than mean-field theory.   For still higher carrier densities
the typical distance between nearby Mn ions will
become larger than the band electron Fermi wavelength, causing the sign
of the typical exchange coupling to oscillate in an RKKY fashion.
We expect the resulting frustration to make the ferromagnetic state
unstable, possibly leading to a regime of spin-glass order. 

As discussed above, different disorder realisations with respect to the ion 
positions give almost identical critical temperatures (see also 
Fig.~\ref{fig2}).
This does not mean, however, that the presence of disorder is unimportant.
Numerical finite-size studies at $T=0$ have shown that in the limit of strong
exchange coupling $J_{pd}$ the presence of disorder increases the spin 
stiffness \cite{SKLM:00,unpub} for the regime of densities and masses 
covered by our calculations.   
This, in turn, enhances the critical temperature since, according to 
analytical estimates from spin-wave theory, $T_{c}$ is approximately 
proportional to the spin stiffness \cite{SKLM:00}.
On the other hand, our Monte Carlo results show that the spin stiffness
enhancement does not 
depend on the concrete disorder realization.

\subsubsection{Local vs. global polarisation}

\begin{figure}
\centerline{\includegraphics[width=8cm]{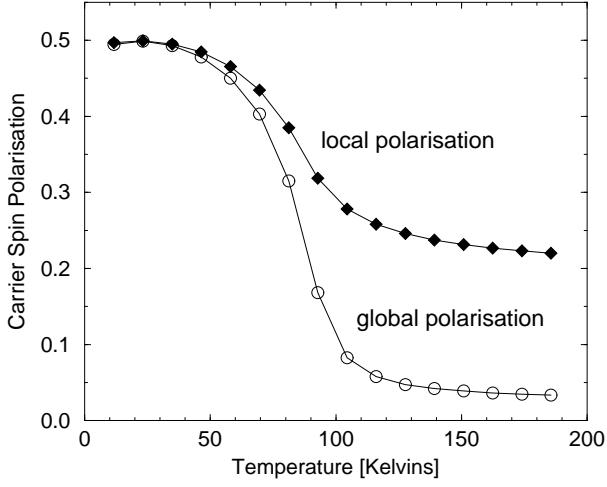}}
\caption{Local versus global carrier spin polarisation for the same
system as in Fig.~\protect{\ref{fig1}} as a function of temperature.
The local carrier polarisation remains finite above $T_{c}$.
\label{fig7}}
\end{figure}

In mean-field theory, the free carrier band spin-splitting 
vanishes as the critical temperature is approached.
This is, however, not generally the correct physical picture.
When long-wavelength collective fluctuations \cite{KLM:00,SKLM:00}, which are 
neglected by mean-field theory, drive the breakdown of ferromagnetism the 
{\it local} carrier spin polarization can remain finite even above $T_c$. 
Ferromagnetism and the {\em global} spin polarization disappears only because 
of the loss of long-range spatial coherence.
The applicability of this picture to the studied parameter ranges of the 
kinetic exchange model is confirmed by our Monte Carlo studies.
In Fig.~\ref{fig7} we compare the global (Eq.~(\ref{carrmag})) and local
carrier spin polarizations, 
\begin{equation}
m_{loc}=\overline{\left\langle\frac{|\vec s(\vec r\,)|}{n(\vec r\,)}
\right\rangle}\,,
\end{equation}
which is the thermally and spatially averaged ratio of the {\em modulus}
of the carrier spin density $\vec s(\vec r\, )$ and the local carrier density 
$n(\vec r\, )$. The system parameter are the same as in Fig.~\ref{fig1}.
While the global spin polarization vanishes at the critical temperature, the
local polarization remains finite and saturates at about fourty percent
of its $T=0$ value.

\subsection{Six-band model}

We now turn to the six-band model in which the kinetic energy part is given by 
the Kohn-Luttinger Hamiltonian (\ref{KL}).
To model GaAs we use $(\gamma_{1},\gamma_{2},\gamma_{3})=(6.85,2.1,2.9)$ for 
the Luttinger parameters, and the spin-orbit coupling energy is 
$\Delta_{so}=0.34{\rm eV}$.

\begin{figure}
\centerline{\includegraphics[width=8cm]{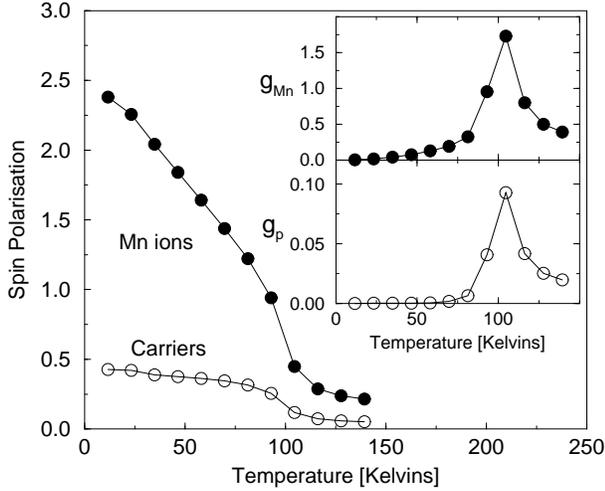}}
\caption{Magnetisation curves for Mn ions and carriers in the six-band model
for n exchange coupling of $J_{pd}=0.15{\rm eVnm}^{3}$. The carrier density is
$n=0.1{\rm nm}^{-3}$ with an Mn ion density of $N=1.0{\rm nm}^{-3}$
in a volume of $V=280{\rm nm}^{3}$. As in the case of parabolic bands,
the ferromagnetic transition in clearly and consistently signalled
by pronounced peaks in the magnetic fluctuations shown in the insets.
\label{fig8}}
\end{figure}

Figure~\ref{fig8} shows typical magnetisation data for a system with
exchange coupling $J_{pd}=0.15{\rm eVnm}^{3}$, carrier density is
$n=0.1{\rm nm}^{-3}$, and Mn ion density $N=1.0{\rm nm}^{-3}$
in a volume of $V=280{\rm nm}^{3}$. The wavevector cutoff
here is $k_{c}=2(2\pi/L)$ offering 192 single-particle states to
the 28 carrier in the system.

As in the case of parabolic bands, a 
ferromagnetic transition is clearly signalled by pronounced peaks 
in the magnetic fluctuations of both Mn ions and carriers.
We find that, in contrast to the parabolic two-band model, the carrier 
magnetisation is already reduced at temperatures well below $T_{c}$.
Another difference compared to the parabolic-band model concerns the shape of 
the magnetic fluctuations for the manganese ions and the carriers
as a function of temperature.
Although both curves indicate the same value for $T_{c}$, their shape in the
vicinity of $T_c$ is slightly different and the ratio of the 
their fluctuations is smaller than $25$ (the square of the ratio of
spin lengths involved).
These differences arise because of the more complicated band structure
and the spin-orbit coupling present in the Kohn-Luttinger Hamiltonian.

As for the two-band model, the data shown in Fig.~\ref{fig8} are the average 
over five different disorder realisations. Also in the present case, the 
dependence of the data on the concrete realisation is very weak.
Moreover, we again find that the local carrier polarization remains finite 
above $T_c$ while the global magnetization vanishes.

\begin{figure}
\centerline{\includegraphics[width=8cm]{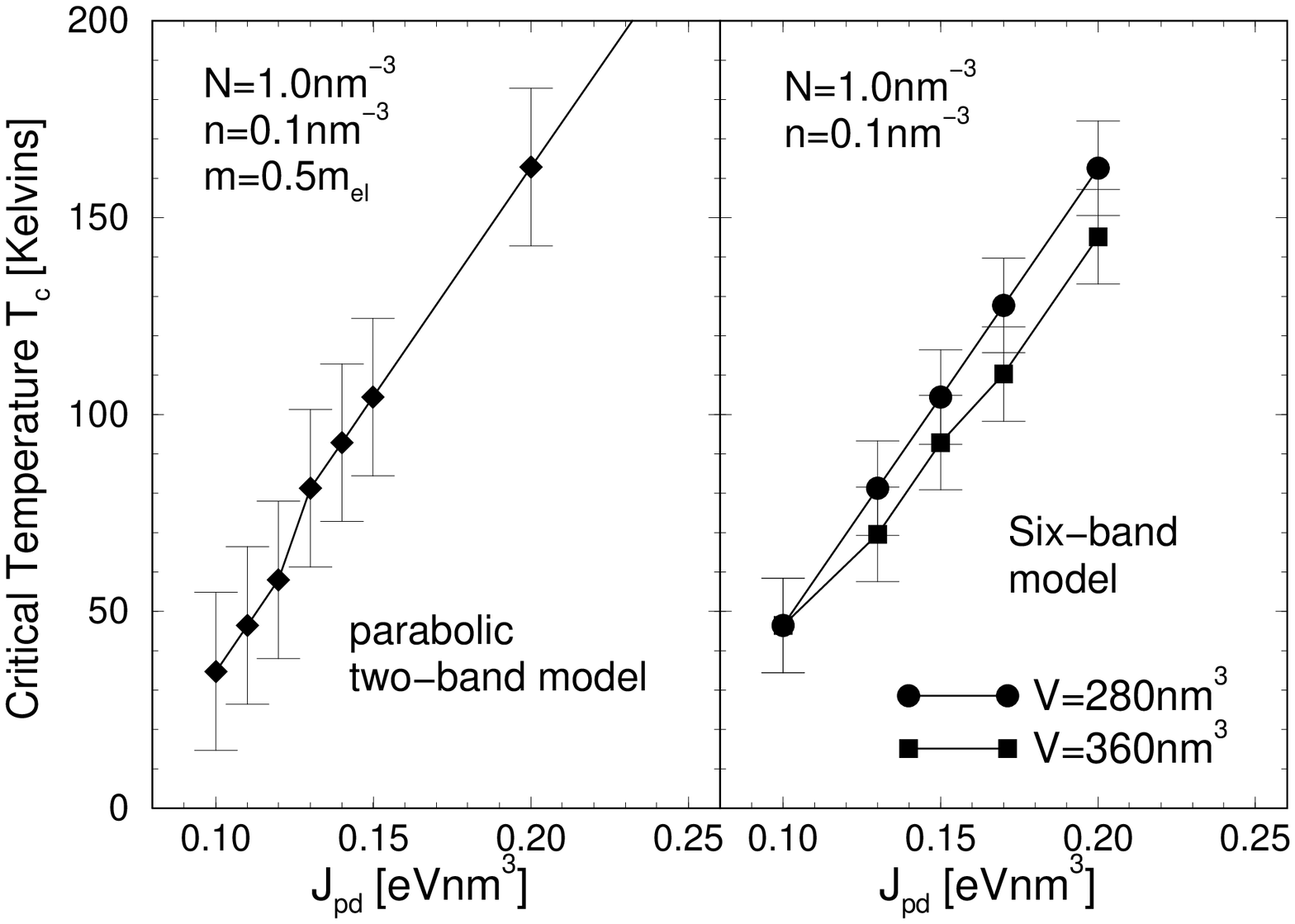}}
\caption{The right panel shows critical temperature $T_{c}$ as a function of 
the exchange parameter $J_{pd}$ for the same particle densities as in
Fig.~\ref{fig8} and two different system sizes. Both data sets agree within
error bars and show a linear dependence of $T_{c}$ on $J_{pd}$.
In the left panel, the corresponding data for the parabolic two-band model
with an effective mass of half the bare electron mass is plotted.
The latter system is a reasonable approximation to the six-band case.
\label{fig9}}
\end{figure}

In the right panel Fig.~\ref{fig9} we plot the transition temperature as a 
function of the exchange coupling $J_{pd}$ for two different system sizes. 
Both data 
sets agree within error bars and show a linear dependence of $T_{c}$ on 
$J_{pd}$.  This finding is the same as for the two-band model and contrasts 
with mean-field theory which predicts $T_c \propto J_{pd}^2$. The left panel
of Fig.~\ref{fig9} shows the transition temperature as a 
function of $J_{pd}$ for the parabolic model for an effective mass
$m^{\ast}=0.5m$. This value is close to the heavy-hole mass in the
Kohn-Luttinger model for parameters appropriate for GaAs as given above.
The data in the left panel can be obtained from the left panel of 
Fig.~\ref{fig6} via the scaling relation (\ref{scalerel}).
Comparing the two panels of Fig.~\ref{fig9} demonstrates that, in the range of 
carrier densities studied, a single parabolic band with an
effective mass close to that of the heavy-hole Kohn-Luttinger-model band 
provides a reasonably good approximation to the behavior of the
six-band system.


\section{Summary}

We have given a detailed report on Monte Carlo studies of a model
that captures essential aspects of ferromagnetism in 
(III,Mn)V ferromagnets.   In the kinetic-exchange model we study 
localized magnetic moments formed by Mn ions are coupled antiferromagnetically 
to carriers (holes) in the valence band.
In most of our simulations we describe the valence-band carriers by simple 
parabolic bands but we have investigated a more realistic 
six-band $\vec k \cdot \vec p$ model as well.
We treat the manganese spins as classical magnetic moments, and take the 
carriers effectively at zero temperature.  Both approximations are 
comparatively weak.
Our treatment accounts for spatial fluctuations in the magnetization and for
disorder induced by the randomness of the Mn ion positions, effects which are
both effects are neglected in a mean-field description.

We outline the Hybrid Monte Carlo algorithm employed and discussed its 
advantages compared to other methods.
A complete account of important technical details of the computation is given. 
We have shown that the effects of different disorder realizations, wavevector 
cutoff, and finite-size effects are well under control.
Our results depend weakly on the regularization parameter $a_0$ which
describes the spatial range of the exchange mechanism between carriers and 
ions.

We find that due to spatial fluctuations and correlations the total 
magnetization and therefore also the ferromagnetic transition temperature $T_c$
is considerably reduced in comparison to mean-field estimates, increasingly
so as the exchange coupling strength and the band effective mass are increased.
In contrast with mean-field-theory predictions, $T_c$ is
not a linear function of the effective band mass and is not proportional to
the square of the exchange coupling constant.
The discrepancy becomes more severe for higher predicted 
mean-field critical temperatures.
It remains an interesting to what extend these deviations from mean field
theory are influenced by the presence and type of disorder
with respect to the manganese positions.

The importance of spatial fluctuations of the magnetization, neglected in a
mean-field description, is emphasized by the fact that the local 
carrier spin polarization remains finite even well above the ferromagnetic 
transition temperature.

\acknowledgements{We thank Glenn Martyna for bringing the advantages of
the Hybrid Monte Carlo algorithm to our attention, and M. Abolfath,
A. Burkov, T. Jungwirth, B. Lee, H.~H. Lin, and Y. Sato 
for useful discussions. J.~S. and J.~K. were supported by
the Deutsche Forschungsgemeinschaft. This research was supported by the
Office of Naval Research
and the Research Foundation of the State University of New York
under grant number N000140010951 and by 
the Indiana 21st Century Fund.}

\begin{thebibliography}{50}

\bibitem{revDMS} J.~K. Furdyna, J. Kossut, {\it Diluted Magnetic 
Semiconductors}, in {\it Semiconductors and Semimetals}, volume 25, 
Academic Press (1988); 
T. Dietl, {\it Diluted Magnetic Semiconductors}, in {\it Handbook of
Semiconductors}, volume 3B, North-Holland (1994).

\bibitem{Ohn:96} H. Ohno, A. Shen, F. Matsukura, A. Oiwa, A. Endo, 
S. Katsumoto, Y. Iye, Appl. Phys. Lett. {\bf 69}, 363 (1996).

\bibitem{Ohn:98} H. Ohno, Science, {\bf 281}, 951 (1998).

\bibitem{Ohn:SSCreview} For a recent review see H. Ohno,
F. Matsukura, Solid State Commun. {\bf 117}, 179 (2001).      

\bibitem{Pri:98} G.~A. Prinz, Science {\bf 282}, 1660  (1998).

\bibitem{dietl_mft}
T. Dietl, A. Haury, Y.~M. d'Aubign{\'e}, Phys. Rev. B {\bf 55}, R3347 (1997); 
T. Dietl, H. Ohno, F. Matsukura, J. Cibert, D. Ferrand, Science {\bf 287}, 
1019 (2000);
T. Dietl, H. Ohno, F. Matsukura, Phys. Rev. B {\bf 63}, 195205 (2001).

\bibitem{Ramin} M. Abolfath, T. Jungwirth, J. Brum, A.~H. MacDonald, 
Phys. Rev. B {\bf 63}, 054418 (2001).

\bibitem{us_mft}
T. Jungwirth, W.~A. Atkinson, B.~H. Lee, A.~H. MacDonald, Phys. Rev.
B {\bf  59},  9818  (1999);
B.~H. Lee, T. Jungwirth, A.~H. MacDonald, Phys. Rev. B {\bf 61}, 15606
(2000); B.H. Lee, Tomas Jungwirth and A.H. MacDonald,
Bull. Am. Phys. Soc. {\bf 46}, 509 (2001).

\bibitem{sham_mft} J. Fernandez-Rossier and L. J. Sham,
Bull. Am. Phys. Soc. {\bf 46}, 509 (2001).

\bibitem{sanvito} S. Sanvito, P. Ordejon, N.~A. Hill, Phys. Rev. B {\bf 63},
165206 (2001);
S. Sanvito, N.~A. Hill, cond-mat/0011372.

\bibitem{butler} T.~C. Schulthess, W.~B. Butler, private communication (2000). 

\bibitem{KLM:00} J. K\"onig, H.~H. Lin, A.~H. MacDonald, Phys. Rev. Lett.
{\bf 84}, 5628 (2000); 
to appear in ''Interacting Electrons in Nanostructures'', edited by R. Haug 
and H. Schoeller (Springer), cond-mat/0010471.  

\bibitem{SKLM:00} J. Schliemann, J. K\"onig, H.~H. Lin, A.~H. MacDonald, 
Appl. Phys. Lett. {\bf 78}, 1550 (2001).

\bibitem{HMC}
S. Duane, A.~D. Kennedy, B.~J. Pendleton, D. Roweth, Phys. Lett. B {\bf 195},
216 (1987).

\bibitem{sakai} O. Sakai, S. Suzuki, K. Nishizawa,
preprint (2000).

\bibitem{Bhatt} X. Wan, R.~N. Bhatt, cond-mat/0009161;
R.~N. Bhatt, M. Berciu, cond-mat/0011319.

\bibitem{Omiya} T. Omiya, F. Matsukura, T. Dietl, Y. Ohno, T. Sakon,
M. Motokawa, and H. Ohno, Physica E {\bf 7}, 976 (2000).

\bibitem{CP} R. Car, M. Parrinello, Phys. Rev. Lett. {\bf 55}, 2471 (1985);
G. Galli, M. Parrinello in {\it Computer Simulations in Materials Science},
M. Meyer, V. Pontikis (eds.), Kluwer (1991); 
P.~E. Bl\"ochl, M. Parrinello, Phys. Rev. B {\bf 45}, 9413 (1992).

\bibitem{Bhatta:00}
A.~K. Bhattacharjee, C. Benoit $\grave { \rm a}$ la Guillaume, Solid State 
Comm. {\bf 113}, 17 (2000).

\bibitem{ChKo:99} W.~W. Chow, S.~W. Koch, {\it Semiconductor-Laser
Fundamentals}, Springer (1999).

\bibitem{LuKo:55} J.~M. Luttinger, W. Kohn, Phys. Rev. {\bf 97}, 869 (1955).

\bibitem{meyer} For a recent review see I. Vurgaftman, J. R.
Meyer, L. R. Ram-Mohan, Appl. Phys. Rev. (in press). 
                                                                    
\bibitem{AFGLM:00}
J.~A. Alonso, L.~A. Fernandez, F. Guinea, V. Laliena, V. Martin-Mayor,
Nucl. Phys. B {\bf 596} 587 (2001).

\bibitem{unpub} J. K\"onig, J. Schliemann, A.~H. MacDonald, unpublished.

\end{thebibliography}
\end{document}